# Magneto-Optical Imaging of Magnetic-Domain Pinning Induced by Chiral Molecules

Yael Kapon[1], Fabian Kammerbauer[2], Shira Yochelis[1], Mathias Kläui[2], Yossi Paltiel[1]

[1] Institute of Applied Physics, Rachel and Selim School of Engineering, The Hebrew University of Jerusalem, Jerusalem 9190401, Israel
[2] Institute of Physics, Johannes Gutenberg University Mainz, Staudingerweg 7, 55128 Mainz, Germany

Abstract:
Chiral molecules have the potential for creating new magnetic devices by locally manipulating the magnetic properties of metallic surfaces. When chiral polypeptides chemisorb onto ferromagnets they can induce magnetization locally by spin exchange interactions. However, direct imaging of surface magnetization changes induced by chiral molecules was not previously realized. Here, we use Magneto-optical Kerr microscopy to image domains in thin films and show that chiral polypeptides strongly pin domains, increasing the coercive field locally. In our study, we also observe a rotation of the easy magnetic axis towards the out-of-plane, depending on the sample's domain size and the adsorption area. These findings show the potential of chiral molecules to control and manipulate magnetization and open new avenues for future research on the relationship between chirality and magnetization.

Introduction:
Over the last thirty years, there has been a growing interest in the spin-selective properties of chiral organic materials[1]. When an electron passes through a chiral molecule, it exhibits a preferred spin polarization corresponding to the molecular chirality, this phenomenon is known as the Chirality Induced Spin Selectivity (CISS)[2,3].

Since its discovery, the CISS effect has shown great promise for applications in spintronics[4] and enantioselective chemistry[1,5]. For both lines of study, the interface between chiral molecules and ferromagnets was studied extensively. Two complementary effects have been utilized for this purpose. The first is the enantioselective adsorption of chiral polypeptides on perpendicularly magnetized samples[6–12]. The second is surface magnetization induced by chiral molecular adsorption[13–16], where the chemisorption of L(D)- α helix polyalanine (AHPA) resulted in local magnetization in the up (down) direction perpendicular to the plane without external stimuli such as light or electric field. Both effects were attributed to thermally induced spin-exchange interactions with the surface. Furthermore, it was found that there is a correlation between the magnetic easy axis and the tilt angle of the molecule[17,18].

The magnetization induced by chiral polypeptides was imaged before using magnetic force microscopy (MFM) imaging[13], measuring the magnetic stray field. Sharma et al.[19] showed the magnetization switching of Co thin layer induced by chiral polypeptides using magneto-optical imaging and spectroscopy. While this method measures the magnetization directly, the local flipping of domains in the vicinity of chiral polypeptides has yet to be imaged.

Here, we used Magneto-Optical Kerr effect (MOKE) microscopy to directly image the impact of chiral polypeptides on the magnetic properties in a sample with sizeable perpendicular magnetic anisotropy and a sample with in-plane aligned magnetic domains.

MOKE microscopy is a non-invasive imaging method that enables direct imaging of surface magnetization with high spatial resolution[20]. This makes it especially interesting to CISS-related research enabling measurements of the chirality-induced magnetization in a direct and fast way. Kerr microscopy is based on the magneto-optic Kerr effect, which occurs when polarized light is reflected off a magnetized surface. The reflected light undergoes a rotation in polarization due to the interaction between the magnetic moment of the surface and the electric field of the incident light. Different magnetization directions on the surface result in different magnitudes of polarization rotation, which leads to contrast differences in the Kerr microscopy image (see Figure 1(a)). This enables direct imaging of magnetic domains and provides valuable insights into the magnetic properties of the sample.

Figure 1(b) presents a sketch of the magnetization changes induced by chiral polypeptides and the domain switching induced by chiral molecule adsorption. The local domain switching can be imaged by MOKE microscopy (Figure 1(b)). A non-magnetized magnetic material will have domains magnetized in all directions with zero net magnetization (Figure 1(b) right). When chiral peptides adsorb on the surface of the ferromagnet, the domains in the vicinity of the chiral molecule are found to align toward the molecular axis of the molecules (Figure 1(b) 2). Following the molecule adsorption, a magnetic field can be applied, aligning the magnetic domains with the field, except the ones in the vicinity of the chiral polypeptides, which will be pinned by the polypeptides (Figure 1(b) 3). This will appear as a changed contrast in the MOKE microscope's image.

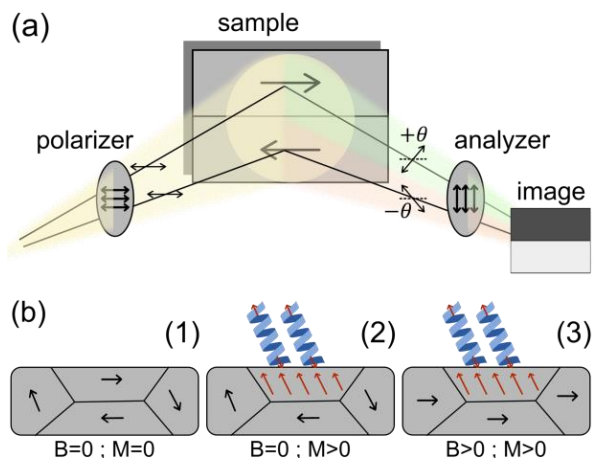

**Figure 1: magnetization of domains by chiral polypeptides.** (a) Magneto-Optical Kerr effect (MOKE) microscopy is used to image the effects of chiral polypeptides on magnetic materials. linearly polarized light, upon reflection from the surface, rotates in different angles due to different directions of magnetization. The different rotations result in different contrasts in the image plane allowing for domain imaging. (b) 1- In the absence of an external magnetic field, a demagnetized material lacks a preferential magnetization direction in its domains. 2- However, the adsorption of chiral polypeptides on the surface magnetizes and aligns the domains near them. 3- Upon applying a magnetic field, the domains align with the field except those near the chiral polypeptides. The polypeptide manipulates the magnetic domains in its vicinity.

Results:

To directly image the magnetization induced by chiral polypeptides we used a commercial Evico-Magnetics GmbH, room temperature, Magneto-optical Kerr effect (MOKE) microscope, equipped with electromagnets in a direction perpendicular and parallel to the sample surface. To enhance the magnetic signal contrast, the sample was saturated using a 2 mT out-of-plane magnetic field, and the saturated image was subtracted from the optical image. Reduced noise was achieved by utilizing active piezo-electric stabilization during the measurement.

It is important to note that the contrast obtained by the MOKE microscope is a qualitative description of the magnetization rather than a quantitative one. The contrast corresponds to a rotation of polarization relative to the analyzer and represents areas of different directions of magnetization relative to the incident plane of light. In the polar mode, we imaged domains perpendicular to the surface plane, while in the longitudinal mode, we imaged in-plane magnetized domains.

Ta(5)/CoFeB(0.9)/MgO(2)/Ta(2)/Au(5nm) thin films with a significant perpendicular magnetic anisotropy (PMA) were prepared using magnon sputtering using a Singulus Rotaris sputter deposition tool. The samples were intentionally designed to enable easy imaging of domains in the MOKE polar configuration at room temperature; a detailed description can be found in the methods section. Magnetization loops are presented in Figure 2(a). To introduce optically detectable molecular aggregates to the surface, an L-α helix polyalanine (L-AHPA) solution was drop-casted onto the sample and allowed to dry, forming visible aggregates as illustrated by blue arrows in Figure 2(b) top. As a reference, a-chiral 11-mercapto undecanoic acid aggregates were prepared in the same way. A detailed description of the drop-casting method can also be found in the methods section. Subsequently, the domains in the proximity of the chiral aggregates, indicated by orange arrows in the bottom panel of Figure 2(b) bottom, were imaged.

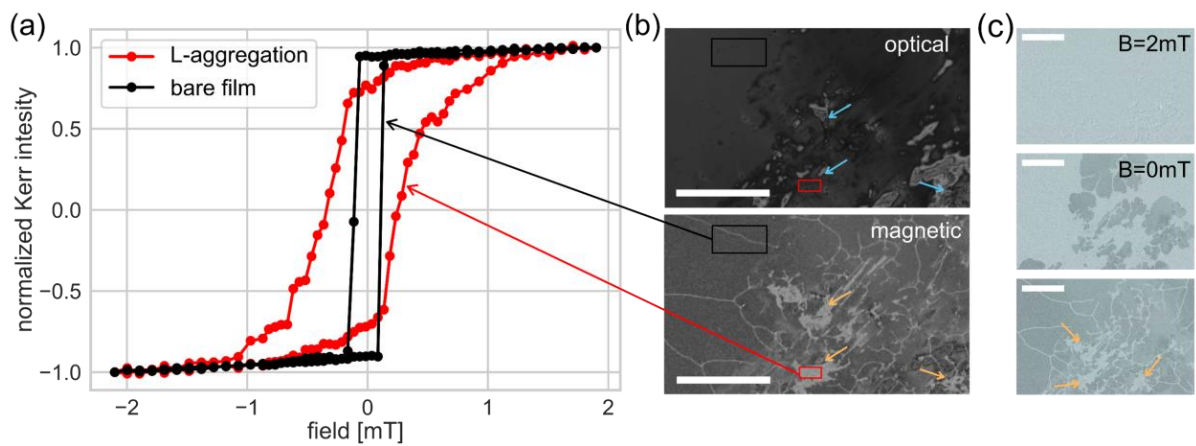

**Figure 2: magnetization induced by chiral polypeptides.** (a) magnetization hysteresis loops were measured for Ta/CoFeB/MgO/Ta/Au thin films around the aggregation of L-α helix polyalanine (red) and at a distance from the aggregation (black) in a polar configuration. The chiral polypeptides increased the coercive field of the sample by 1 mT, which is ten times the original value. (b) Top - an optical image of the chiral aggregates (blue arrows). Bottom - a MOKE image of the domains held by the chiral aggregation (orange arrows). (c) demonstration of domain spreading from the chiral polypeptides. At 2 mT, the magnetization is uniformly oriented. When the field is turned off, the magnetization changes and oppositely magnetized domains appear. At a steady state, domains are fixed by the chiral aggregates. The scale bar is 100 μm.

Figure 2(a) displays the magnetization induced by chiral polypeptides. The domains of Ta/CoFeB/MgO/Ta/Au thin films were imaged while sweeping the magnetic field direction from -2mT to +2mT. The Kerr intensity was averaged over selected areas (red/black squares). This averaged intensity corresponds to the surface magnetization, which is normalized relative to the saturated intensity. Magnetization loops for Ta/CoFeB/MgO/Ta/Au thin films were measured around the L-α helix polyalanine aggregation (red area), and at a distance from the aggregation (black area), in a polar configuration. A clear change in the magnetization loops was observed near (red) and away from (black) the aggregation.

Since the coercive field of the continuous film of the untreated samples is around 0.1 mT, most domains of the sample flip and align with the field direction as visible by the contrast change to dark when the field is swept from negative to positive and passes a value of 0.1 mT. This expected behavior is not seen for the domains near the aggregation which remain with bright contrast. It takes another 1 mT to flip all the domains close to the aggregation. Reversing the field direction, sweeping positive to negative, results in the opposite effect. At -0.1mT, most domains flip to become bright, but it takes another 1 mT in the negative direction to flip the domains close to the aggregation. The effect of the chiral polypeptides is observable in different areas close to the aggregation and is reproducible. Supplementary video S1 shows an example of flipping as the field is swept from negative to positive and back. This indicates that the adsorbed layer has drastically impacted the anisotropy of the magnetic layer.

The chiral polypeptides increase the coercive field of the chosen PMA sample by 1 mT, which is ten times the original value. Supplementary Figure S1 presents the same experiment with a-chiral aggregates were no change in the magnetic properties of the film was recorded, indicating the change is derived from the chirality and not from any mechanical or chemical changes to the surface. In addition, the solvent itself does not change the magnetic properties of the magnetic layer as seen in supplementary Figure S2. Previous studies [13] have reported chirality-induced magnetization using MFM, however, direct imaging of the surface magnetization itself had not been achieved before. It is worth noting that the shape of the magnetization loop changes as well, attributed to the sample area chosen and the extent of flipping. The closer the domains are to the aggregation, the magnetization changes more abruptly. In addition, the oxide layer in the chosen PMA sample likely screens the effect of the chiral polypeptides, and metallic thin films without oxidation may present a stronger effect.

To investigate the effect of the chiral polypeptides in a thermally active dynamic regime, we first saturated the magnetic sample and then turned off the magnetic field to record the domain flipping in real-time, as shown in supplementary video S2. The corresponding images from the video are displayed in Figure 2(c). At 2 mT, the magnetization is uniform, and the domains appear bright (Figure 2(c) top). When the magnetic field is turned off, the magnetization flips and dark domains emerge (Figure 2(c) middle). The dark domains spread from the chiral aggregates, indicating that the chiral aggregates act as pinning sites. In a steady state, the chiral aggregates pin the light domains in place, as indicated by the orange arrows in Figure 2(c) bottom. This observation corroborates the effect seen in the magnetization hysteresis loops. In the a-chiral case, no correlation was seen between the domain flipping and aggregates, as presented in Figure S1.

The induced changes of the magnetization were observed perpendicular to the plane and using aggregates that do not have a specific orientation compared to the sample. Previous papers have also shown a correspondence between the magnetic easy axis and monolayer tilt angle[17,18]. These studies characterized the magnetic easy axis fully by vectorial magnetic field measurements while measuring the monolayer tilt angle using atomic force microscopy and found a correlation between the two. To observe the correlation, a sample with in-plane domains with a monolayer adsorption was imaged. A sample of Ti(2)/Ni(80)/Au(5nm) was prepared by magnon sputtering as described in the methods section. The resulting sample had in-plane domains that were imaged in the longitudinal mode. The sample was prepared in such a way that the averaged domain size was on a 50 µm scale, large enough to be imaged optically (in the longitudinal configuration).

On the Ni sample, we adsorbed a monolayer of L-AHPA using a self-assembly process, discussed further in the methods. Previous studies found such monolayers have a uniform tilt angle of 30-60 degrees normal to the surface[18,21] due to their self-assembling properties and so it is well-suited for exploring the effects of the molecular axis versus the magnetic easy axis. By utilizing electron beam lithography, we were able to selectively adsorb polypeptides onto different rectangular areas with widths starting from 5 µm up to 60 µm, resulting in varying sizes of adsorption regions that enable to probe different local effects.
s
Figure 3 shows magnetization loops in the longitudinal configuration of L-AHPA adsorbed on Ti(2)/Ni(80)/Au(5) samples of various monolayer sizes (ranging from 5*300 µm² to 60*1000 µm²). The remanent magnetization (the magnetization at zero field) for the largest monolayer area (red) is smaller than that of the bare magnetic sample (purple), indicating a rotation of the easy axis from the in-plane toward the out-of-plane. The tilt of the easy axis depends on the monolayer area, with the hysteresis loop softening as the monolayer size increases, reaching saturation above 20*600 µm² absorption area, which matches the sample's domain size (50*10µm²).  The rotation of the easy axis is connected to the CISS effect due to the correlation between the tilt angle of the monolayer's molecular axis compared to the surface as sketched in Figure 3(b).

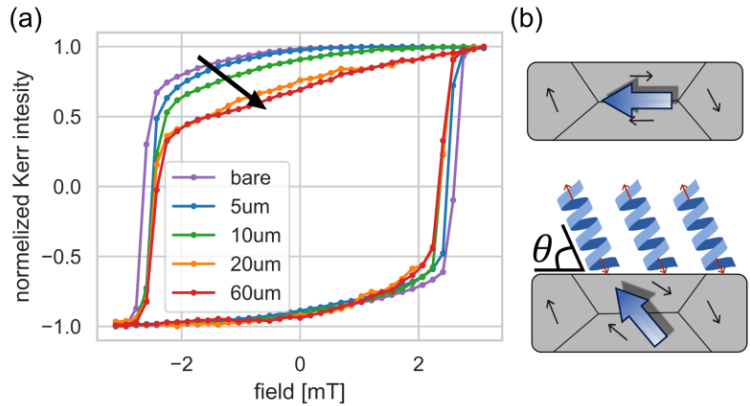

Figure 3: rotation of the easy magnetic axis by a chiral monolayer in an in-plane ferromagnet. (a) In-plane magneto-optical magnetization loop for a Ti(2nm)/Ni(80nm)/Au(5nm) sample (purple) with L-α helix polyalanine adsorbed on different areas: 60*1000 µm²(red), 20*600 µm² (orange), 10*300 µm² (green), and 5*300 µm² (blue). As the adsorption area increases, the remanent magnetization for the in-plane contrast measurement decreases, indicating a rotation of the easy axis towards the out-of-plane direction. (b) illustration of the rotation of the magnetic easy axis (blue). After monolayer adsorption, the magnetic easy axis couples with the molecular tilt angle ($\theta$).

To test the importance of domain size compared to the adsorption area, two samples with different domain sizes were used. Thus, the remanent magnetization in the MOKE magnetization loops on samples of Ti(2nm)/Ni(30/80nm)/Au(5nm) sputtered and patterned with a chiral L-AHPA monolayer by a self-assembly process was measured. Figure 4a shows the remanent magnetization of the sample decrease with adsorption area, with both 8nm/80nm/5nm (red) and 8nm/30nm/5nm (blue) samples showing a decrease, but with a smaller effect in the latter. The different Ni thickness translates to different average domain sizes as seen in Figure 4(b). While the 80 nm Ni thickness has smaller domains (50 µm) the 30 nm Ni domains are larger, above the optical field of view of the microscope (>500 µm). Therefore, the results point to the importance of domain size as compared to the adsorption area. There is an expected match between the domain size and adsorption area.

Lastly, to connect all the above results to the molecular chirality in opposition to surface effects induced by adsorption or treatment prior to adsorption, we tested the magnetization loops of Ti(2nm)/Ni(30)/Au(5nm) (orange) with a patterned monolayer of L- (purple), size was 60*1000 µm$^2$, D- (red) AHPA, and a-chiral mercaptoundecanoic acid (green), size was 60*700 µm$^2$. The rotation of the easy-axis was observed similarly between L and D chirality, while a-chiral mercaptoundecanoic acid adsorption had only a small effect. This indicates the changes in magneto-optical properties observed in this paper were a result of the chiral molecule-induced manipulation of the magnetic properties.

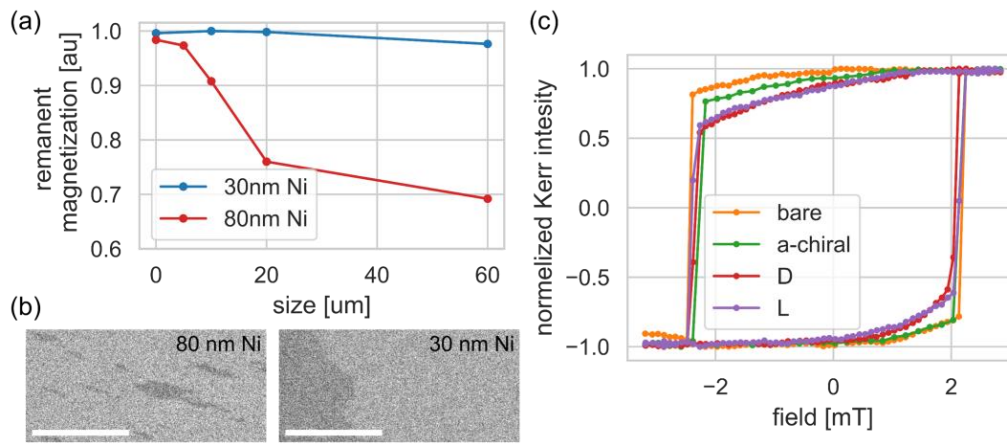

Figure 4: Effect of L-α helix polyalanine adsorption on magneto-optical effects of Ti(2nm)/Ni(30/80nm)/Au(5nm) samples. (a) The remanent magnetization of the sample decreases with adsorption area, with both 8nm/80nm/5nm (red) and 8nm/30nm/5nm (blue) samples showing a decrease, but with a smaller effect in the latter due to its larger domain size. (b) magnetic imaging shows smaller domains for 80nm Ni (50*10 µm$^2$) than 30nm Ni (>500*500 µm$^2$). (c) magnetization loops of L- (purple) and D- (red) α helix polyalanine adsorption show a rotation of the easy axis (adsorption size is 60*700 µm$^2$) while a-chiral mercaptoundecanoic acid (green) adsorption has only a small effect compared with the Ti(2)/Ni(30)/Au(5) sample (orange). The scale bar is 100 µm.

Discussion:
In this study, we presented a direct spatial imaging of chiral-molecule-induced enhanced coercivity (pinning) of magnetization due to manipulation of the anisotropy. Our results demonstrate that chiral aggregates caused a drastic increase in the coercive field of 1mT in a perpendicular magnetization of thin films. This change in magnetic properties was accompanied by the strong pinning of domains to the chiral aggregates, as revealed by domain imaging. In addition, real-time measured domain flipping originating from the chiral aggregates was observed, suggesting that there is an interacting force that is stable after adsorption between the magnetic domains and the chiral molecules. Different mechanisms can lead to such behavior. In general, the observed changes of the magnetic switching imply a strongly changed effective anisotropy. We relate this to a possible interaction between thermally induced spin polarization[15,22,23], and the spins of the magnetic layer. Several theoretical papers[11,24,25] and experimental observations[13,26–29] suggest the induced change in magnetic properties is derived from spin-exchange interactions between the chiral polypeptides and the sample. While magnetization induced by chiral polypeptides had been previously observed using MFM[13], this study has achieved direct imaging of surface magnetization. The interplay between molecular tilt angle and magnetic easy axis, which was previously explored[17,18], was confirmed indicatively through direct imaging of the domains in the vicinity of the chiral monolayer. The main observation in this work may also be relevant to the origin of homochirality in life as suggested by the avalanche mechanism presented in by Ozturk et al[30].

Samples with observable domains in the longitudinal configuration and in-plane easy axis were fabricated to observe the effects parallel to the surface. In these samples, a direct coupling between the molecular alignment axis of the chiral monolayer and the magnetic easy axis was measured, as evidenced by a decrease in the remanent magnetization of the magnetic layer after the chiral monolayer's adsorption. We also investigated the connection between domain size and monolayer area and found that the maximum effect is achieved when there is a matching between the two. A reference sample showed that while L and D AHPA reduced the remanent magnetization of the magnetic layer, the a-chiral monolayer had little effect on the magnetic properties.

As demonstrated in this study, MOKE microscopy and MOKE magnetometry offer promising avenues for observing and studying induced pinning effects by chiral molecules. The ability to directly observe surface magnetization is particularly important in CISS-related studies, where the effects are strongest at the interface. The magnetization effect observed in our study has the potential to be scaled down to a sample with smaller domains and adsorption areas, as demonstrated in previous work, using different imaging methods[13]. It is important to note that the change in coercivity leading to domain-pinning observed in our study is a local and permanent change in magnetic properties and will change the dynamic behavior. An interesting approach would be to observe the domains in real-time during the adsorption process to directly observe the magnetization induced by the adsorption itself, which however goes beyond the scope of the current work.

Another avenue of research that could be explored using the same measurement concepts demonstrated in this study is the temperature dependence of the chirality-induced magnetization effect. Both the CISS effect and magnetization of ferromagnets are known to be temperature-dependent[15,22], and studying the effect of temperature on these measurements would provide valuable insights into the underlying mechanisms of the effect. These studies would further our understanding of the complex interplay between chirality and magnetism, with potential implications for the development of new technologies.

Conclusions:
In this study, we used MOKE microscopy to directly image the changes in the magnetic properties leading to strong pinning of domains in ferromagnetic perpendicular magnetic anisotropy thin films by chiral polypeptides. In addition, tilting of the magnetic easy magnetization axis was observed, towards the molecule alignment axis of an adsorbed molecular monolayer. This non-invasive imaging method allowed us to observe the local magnetization and magnetic effects at the interface between the chiral polypeptides and magnetic samples. Our results show that magnetic domains are affected even outside the adsorption area, while maximum effect is achieved when the domain size matches the adsorption area. Our findings offer new perspectives for the study of the CISS effect and its potential applications in the future.

Methods:

Magnetic sample preparation:
The Ta(5)/CoFeB(0.9)/MgO(2)/Ta(2)/Au(5nm) thin films were deposited by magnetron sputtering using a Singulus Rotaris sputtering system on thermally oxidized silicon substrates. The base pressure in the chamber is around $10^{-8}$ mbar. Films were deposited in Argon atmosphere of around $10^{-3}$ mbar. The film thickness is estimated from sputter rates determined by x-ray reflectivity measurements.

The Ti(2)/Ni(30/80)/Au(5nm) thin films were deposited by magnetron sputtering using AJA, ATC Polaris Series on a Si substrate in an ultrahigh vacuum deposition chamber with a base pressure $5 \cdot 10^{-10}$ Torr. The deposition was done in $3 \cdot 10^{-3}$ Torr with 20 W intensity for Au and 125 W intensity for Ti and Ni layers. Layers were grown in the same chamber without breaking the vacuum.

Materials:
36 - L/D α-helix polyalanine (L/D-AHPA) [[H]-CAAAAKAAAAKAAAAKAAAAKAAAAKAAAAKAAAAK-[OH]] molecules (C stands for cysteine, A for alanine, and K for lysine), as well as a-chiral 11-mercapto undecanoic acid molecules were manufactured by Sigma–Aldrich. A 1mM solution was prepared in ethanol and used in the experiments.

Patterned Monolayer:
Sample cleaning was performed with boiling acetone for 10 minutes followed by boiling isopropanol for 10 minutes and subsequently water. The samples were dried by a Nitrogen gun and then put on a hot plate at 80ºC to evaporate any residual solutions.
Selective adsorption of L/D α-helix polyalanine (L/D-AHPA), or a-chiral 11-mercapto undecanoic was obtained by defining rectangular structures into a poly(methyl methacrylate) (PMMA) resist layer using e-beam lithography, where the Ti/Ni/Au surface was exposed. In order to prepare the exposed Au surface for the adsorption, it was treated by Ozonator to oxidize the surface organic contaminations and then soaked in ethanol for 20 minutes to reduce the produced oxides. Afterwards the molecules were chemically adsorbed trough their tiol end onto the opened areas via an overnight soak in 1 mM solution of the molecules in ethanol in a nitrogen environment[13,18]. The imaging process was performed through the PMMA resist layer, which was not removed after patterning.

Drop casted aggregates:
To create optically visible aggregates, a 5 µL drop of 1 mM L-AHPA or 11-mercapto undecanoic acid solution was drop-casted onto a thin film stack consisting of Ta/CoFeB/MgO/Ta/Au. The drop was allowed to dry completely, and the resulting aggregates were located by scanning the sample using an optical microscope.

Magneto-optical imaging

MOKE imaging and magnetometry were performed by a commercial Evico Magnetics GmbH magnetooptical Kerr microscope. The measurements were taken in the polar (Figure 2) or longitudinal (Figure 3&4) configuration. An in-plane and out-of-plane magnetic field was generated by electromagnets obtained from the microscope supplier powered by two separate Kepco BOP 100-4DL power supplies. For the optical imaging of magnetic sample, 20X or 50X commercial Zeiss objective lenses were used. For the MOKE imaging, first, a magnetic field was applied to saturate the sample. Then, the saturated image was subtracted from the optical image to enhance the magnetic image contrast. Mechanical vibrations of the sample were actively stabilized by a piezo stage. For the Kerr hysteresis measurements, the optical intensity was averaged over a chosen area and then normalized to saturation magnetization.

Acknowledgments:
The authors are grateful to the Zeiss Foundation HYMMS project, and the Deutsche Forschungsgemeinschaft (DFG, German Research Foundation) 403502522-SPP 2137 Skyrmionics, SFB TRR 173 Spin+X (project A01 #268565370 and project B02 #268565370). M.K. acknowledges the support from the European Union's Horizon 2020 research and innovation programme under grant no. 856538 (3D MAGIC) and the TopDyn Center for Dynamics and Topology. Y.P. acknowledges the support of the BSF Grant 2022503, Horizon EU Marie Sklodowska Curie (DN) CISSE grant, and The Ministry of Innovation, Science and Technology grant 7500157.

Autor declaration:
Conflict of interest –
The authors have no conflicts to disclose.
Author Contributions –
**Yael Kapon**: Conceptualization, Investigation, Original Draft Preparation, Review & Editing **Fabian Kammembauer**: Resources, Review & Editing **Shira Yochelis**: Supervision, Review & Editing **Mathias Klaui**: Resources, Supervision, Review & Editing **Yossi Paltiel**: Conceptualization, Supervision, Review & Editing
Data Availability –
The data that support the findings of this study are available from the corresponding author upon reasonable request.

Supplementary material:
The supplementary material contains:
Video S1 – magnetization flipping during magnetization loops.
Video S2 – magnetization flipping when turning off the magnetic field.
Figure S1 – magnetization loops of a-chiral aggregation.
Figure S2 – magnetization loops of ethanol solution vs bare substrate.